\documentclass[showkeys,nofootinbib]{revtex4}

\usepackage{amsfonts}
\usepackage{amssymb}
\usepackage{natbib}
\usepackage[latin1]{inputenc}
\usepackage[babel]{csquotes}
\usepackage{graphicx}
\usepackage[english]{babel}

\usepackage[T1]{fontenc}
\usepackage[centertags]{amsmath}

\usepackage{mathrsfs}
\usepackage{amsmath}
\usepackage{amsthm}
\usepackage{amssymb}
\usepackage{mathtools}
\usepackage{wasysym}
\usepackage{subfigure}
\usepackage{multirow}
\DeclareFontFamily{U}{mathb}{\hyphenchar\font45}
\DeclareFontShape{U}{mathb}{m}{n}{
      <5> <6> <7> <8> <9> <10> gen * mathb
      <10.95> mathb10 <12> <14.4> <17.28> <20.74> <24.88> mathb12
      }{}
\DeclareSymbolFont{mathb}{U}{mathb}{m}{n}

\DeclareMathSymbol{\Sun}{3}{mathb}{"40}

\begin{document}
\title{Semi-analytical and numerical solutions to Teukolsky equations for large fermion mass over black hole mass ratio}
\author{Mattia Villani}
\affiliation{DISPEA, Universit\`a di Urbino Carlo Bo via Santa Chiara, 27 61029, Urbino, Italy}
\affiliation{INFN - Sezione di Firenze via B.Rossi, 1 50019, Sesto Fiorentino, Florence, Italy}
\email{mattia.villani@uniurb.it}
\date{\today}

\begin{abstract}
In a recent paper, we have studied the Teukolsky equations for fermions with mass $m_e\neq 0$ and rotating black hole of mass $M$. There, we have studied two cases: $\tilde{m}_e=m_e\,M^{-1}\ll 1$ and $a\omega\ll 1$; $\tilde{m}_e\ll 1$ and $a\omega \gtrsim 1$. Here we study the two remaining case case in which $\tilde{m}_e\gtrsim 1$ and $a\omega\ll 1$ using a semi-analytical approach and $\tilde{m}_e\gtrsim 1$ and $a\omega\gtrsim 1$ using a numerical approach. This case could be of some interest for the study of the interactions of fermions with small black holes, such as those formed in the last stages of the the Hawking evaporation process.
\end{abstract}
\maketitle

\section{Introduction}
Solutions to the Teukolsky equation for semi-integer spin with zero mass are known, see the review \cite{lrr}, the book \cite{libro} and references \cite{Ferm,SD}, for example. In a recent paper, \cite{mio}, we have tackeld the problem to solve these equations in the non-null mass case, $m_e\neq 0$. We have identified four different regimes depending on the relative largeness of the parameters $\tilde{m}_e=m_e\,M^{-1}$ ($M$ is the black hole mass) and $a\omega$:
\begin{enumerate}
\item Regime $\tilde{m}_e\ll 1$ and $a\omega\ll 1$: in the previous work, we have treated this case in an approximate way finding analytical corrections to the separation constant $\lambda$ and the renormalized angular momentum $\nu$;
\item Regime $\tilde{m}_e\ll 1$ and $a\omega\gtrsim 1$: we have treated this case in a semi-analytical way by finding analytical corrections to the numerical values of $\lambda$ and $\nu$ calculated numerically with usual methods; here we treat the two remaining cases:
\item Regime $\tilde{m}_e\gtrsim 1$ and $a\omega\ll 1$;
\item Regime $\tilde{m}_e\gtrsim 1$ and $a\omega\gtrsim 1$.
\end{enumerate}
We shall start with the last, in order to describe a method inspired to the spectral method of \cite{huges} devised in order to find numerical values of $\lambda$ and $\nu$ for any value of the parameters, in particular for the case in which $a\omega=0$. The third case shall be treated in a semi-analytical way similarly to the second regime using the result obtained from the above method.

The second regime is of course the most important from the astrophysical point of view, since in every known case $M\gg m_e$, but the cases treated here, with large $m_e$ and also large $\tilde{m}_e$ can also be interesting, at least from the theoretical point of view, since they migh be used to shed ligh on the interaction of very small black holes with fundamental particles and fields. Such small black holes might arise from the last stages of the Hawking evaporation process, for example.

The paper is organized as follows: in Section \ref{sec:teu} we introduce the Teukolsky equations; in Section \ref{sec:separation}, we present our method for the numerical calculation of the separation constant; in Section \ref{sec:radial}, we reat the radial Teukolsky equation and in Section \ref{sec:concl} we conclude our treatment. There is also an appendix, Appendix \ref{app:uno}, in which we list the coefficients of the recurrence relation we have obtained during the treatment of the radial equation.

\section{Teukolsky equations for massive fermion}
\label{sec:teu}

The radial Teukolsky equation for a massive fermion in the Kerr spacetime is given by \cite{libro}:
\begin{equation}\label{eq:radial}
\begin{split}
&\Delta^{-s}\frac{d}{dr} \left( \Delta^{s+1} \frac{d{}_sR}{dr}\right) + \left( \frac{K^2-2i s(r-M)K}{\Delta} +4 i s \omega r - \lambda \right){}_sR+\\
&-\left[ \left( \frac{2i s M \tilde{m}_e \Delta}{\sqrt{\lambda}-2i s M \tilde{m}_e r} \right) \left( \frac{d}{dr} -\frac{2i s K}{\Delta}+\frac{(2s+1)(r-M)}{2\Delta} \right) -M \tilde{m}_e^2r^2 \right]{}_sR=0
\end{split}
\end{equation}
while the radial equation is \cite{libro}:
\begin{equation}\label{eq:angular}
\begin{split}
&\frac{1}{\sin(\theta)} \frac{d}{d\theta}\left( \sin(\theta)\frac{d}{d\theta} {}_sS_l^m \right)+\\
+&\left( a^2\omega^2\cos^2(\theta)-\frac{m^2}{\sin^2(\theta)}-2a\omega s \cos(\theta)-\frac{2m s \cos(\theta)}{\sin^2(\theta)}-s^2\cot^2(\theta)+\lambda -a^2\omega^2+2a m \omega+s \right) {}_sS_m^l+\\
&\left( \frac{a M \tilde{m}_e\sin(\theta)}{\sqrt{\lambda}-2am_e s \cos(\theta)} \right) \left( \frac{d}{d\theta} - 2 a m \omega s \sin(\theta)+\frac{2m s}{\sin(\theta)} - a^2 M \tilde{m}_e^2 \cos^2(\theta) \right) {}_sS_m^l=0,
\end{split}
\end{equation}
In the above formulas, we have:
\begin{equation}
\Delta=r^2-2 M r +a^2 \quad K=(r^2+a^2)\omega-a m \quad \tilde{m}_e=\frac{m_e}{M},
\end{equation}
while $m_e$ is the fermion mass and $\lambda$ is a separation constant, $a$ and $M$ are the black hole angular momentum and mass and $\omega$ is the frequency of the incoming radiation.

The presence of the mass term complicates the solution of the equations, since in this case $\lambda$ cannot be found simply by solving an eigenvalue problem as done usually for a massless fermion or a spin 0, 1 or 2 fields. As stated in the introduction, we can recognize four different regimes according to the values of $a\omega$ and $\tilde{m}_e$:
\begin{enumerate}
\item Both $a\omega$ and $\tilde{m}_e$ are \emph{small}: this case can be treated pertubatively in a similar way to what is usually done for other values of the spin, but, in this case, there will be two expansion parameters;\label{itm:small}
\item Both $a\omega$ and $\tilde{m}_e$ are \emph{large}: this requires a full-fledged numerical approach;\label{itm:large}
\item In the third regime, we have that $a\omega\gtrsim 1$, but still $\tilde{m}_e\ll 1$: in this case, the smallness of the fermion mass allows for a perturbative semi-analytical approach, i.e. we can obtain analytical corrections due to the presence of $m_e$ and add them to the numerical solution of the Teukolsy equations for a finite value of $a\omega$, obtained for example, from the spectral method \cite{huges1,huges2};\label{itm:smallarge}
\item In the fourth regime, $a\omega$ is small and $\tilde{m}_e\gg 1$. In this case, the black hole mass $M$ is smaller than the fermion mass $m_e$. This regime must also be treated in a semi-analytical way, similarly the third regime, starting from the numerical solution of the equations for $a\omega=0$ and $\tilde{m}_e\neq 0$, and then applying analytical corrections due to the finiteness of $a\omega$.\label{itm:largesmall}
\end{enumerate}

In this paper, we tackle the problems of regimes \ref{itm:large} and \ref{itm:largesmall}. Here we first have to consider the case with $a\omega=0$, in order to find the separation constant $\lambda$. We start in the following section by rewriting the angular equation in a more suibable form.

\section{Calculation of the separation constant}
\label{sec:separation}

We were nnot able to devise a method for the analytical calculation of the separation constant, so we have used a numerical method similar to the spectral method. We now describe our algorithm.

We first start with the case $a\omega=0$, $m_e\neq 0$ and subtract from the angular equation the spin-weighted spherical harmonics equation, thus getting, for negative spin:
\begin{equation}
\left[\left(\sqrt{\lambda} + a me \cos(\theta)\right)  \left(\lambda - l (1 + l) - \dfrac{1}{4} \right) - a m_e m - a^3 m_e^3 \cos(\theta)^2 \sin(\theta) \right] \; {}_sS_{lm}(\theta) + a m_e \sin(\theta) \dfrac{d\;{}_sS_{lm}(\theta)}{d\theta}
\end{equation}
and for positive spin:
\begin{equation}
\left[\left(\sqrt{\lambda+1} - a m_e \cos{\theta}\right)  \left(\lambda - l (1 + l) + \dfrac{3}{4}\right) + a m_e m - a^3 m_e^3 \cos(\theta)^2 \sin(\theta)\right]\; {}_sS_{lm}(\theta) + a m_e \sin(\theta) \dfrac{d\;{}_sS_{lm}(\theta)}{d\theta}.
\end{equation}

We now multiply on the left by ${}_sY^{*}_{l_1 m_1}(\theta)$ and write:
\begin{equation}\label{eq:dev}
{}_sS_{lm}(\theta) = \sum_{l} A_{l}\;{}_sY_{l m}(\theta).
\end{equation}
We can now integrate the resulting equations thus getting a non-linear system in $A_l$ and $\lambda$. This system is in principle infinite, since one should sum an infinite number spin-weighted spherical harmonics in \eqref{eq:dev}, but we assume we can truncate this expansion up to a value $l_{max}$ since for $l_1$ too far from $l$ the contribution to the sum should be small. However, we have a problem, i.e. there are too many variables: there are $l_{max}+1$ $A_l$ and $\lambda$, while there are only $l_{max}$ equations, so we add the condition that:
\begin{equation}
\sum_l\, A_l^2 = 1
\end{equation} 
so we can hope to solve the system using as initial conditions:
\begin{equation}
A_l = \delta_{l\,l_1} \qquad \lambda = l_1(l_1+1) - s(s+1).
\end{equation}

We can repeat the above procedure for the case $a\omega\neq 0$, thus getting the equations
\begin{equation}
\begin{split}
&\left[\left(\sqrt{\lambda} + a me \cos(\theta)\right)  \left(\lambda - l (1 + l) - \dfrac{1}{4} + 2a\omega m + a\omega\,\cos(\theta)-a^2\omega^2\sin^2(\theta)\right) - a m_e m - a^3 m_e^3 \cos(\theta)^2 \sin(\theta) \right] \; {}_sS_{lm}(\theta) +\\
&+ a m_e \sin(\theta) \dfrac{d\;{}_sS_{lm}(\theta)}{d\theta}
\end{split}
\end{equation}
for negative spin, while for positive spin we have:
\begin{equation}
\begin{split}
&\left[\left(\sqrt{\lambda+1} - a m_e \cos{\theta}\right)  \left(\lambda - l (1 + l) + \dfrac{3}{4} + 2a\omega m - a\omega\,\cos(\theta)-a^2\omega^2\sin^2(\theta) \right) + a m_e m - a^3 m_e^3 \cos(\theta)^2 \sin(\theta)\right]\; {}_sS_{lm}(\theta) +\\
&+ a m_e \sin(\theta) \dfrac{d\;{}_sS_{lm}(\theta)}{d\theta}.
\end{split}
\end{equation}

We reposrt in table \ref{tab:sep_con} the values we have obtained for the separation constant for $m_e=1$ and $a=\{0.1,0.5,0.9,0.9999\}$ for negative spin, $m=-\dfrac{1}{2}$ and $a\omega=\{0,10^{-3}\}$ as an example.

\begin{table}
\centering
\begin{tabular}{ccccccc}
\multicolumn{7}{c}{$m_e=1, a=0.1, a\omega=0$}\\
\hline
&$l=\dfrac{1}{2}$ & $l=\dfrac{3}{2}$ & $l=\dfrac{5}{2}$ & $l=\dfrac{7}{2}$ & $l=\dfrac{9}{2}$ & $l=\dfrac{11}{2}$\\
\hline
$\lambda$ & 0.911728 & 3.97146 & 8.98232 & 15.9871 & 24.9897 & 35.9488\\
\hline
\multicolumn{7}{c}{$m_e=1, a=0.5, a\omega=0$}\\
\hline
&$l=\dfrac{1}{2}$ & $l=\dfrac{3}{2}$ & $l=\dfrac{5}{2}$ & $l=\dfrac{7}{2}$ & $l=\dfrac{9}{2}$ & $l=\dfrac{11}{2}$\\
\hline
$\lambda$ & 0.489338 & 3.86083 & 8.91806 & 15.9399 & 24.9495 & 35.7372\\
\hline
\multicolumn{7}{c}{$m_e=1, a=0.9, a\omega=0$}\\
\hline
&$l=\dfrac{1}{2}$ & $l=\dfrac{3}{2}$ & $l=\dfrac{5}{2}$ & $l=\dfrac{7}{2}$ & $l=\dfrac{9}{2}$ & $l=\dfrac{11}{2}$\\
\hline
$\lambda$ & 0.206599 & 3.78286 & 8.8886 & 15.9143 & 24.9208 & 35.524\\
\hline
\multicolumn{7}{c}{$m_e=1, a=0.9999, a\omega=0$}\\
\hline
&$l=\dfrac{1}{2}$ & $l=\dfrac{3}{2}$ & $l=\dfrac{5}{2}$ & $l=\dfrac{7}{2}$ & $l=\dfrac{9}{2}$ & $l=\dfrac{11}{2}$\\
\hline
$\lambda$ & 0.780308 & 3.77441 & 8.89131 & 15.9136 & 24.9168 & 35.4717\\
\hline
\hline
\multicolumn{7}{c}{$m_e=1, a=0.1, a\omega=10^{-3}$}\\
\hline
&$l=\dfrac{1}{2}$ & $l=\dfrac{3}{2}$ & $l=\dfrac{5}{2}$ & $l=\dfrac{7}{2}$ & $l=\dfrac{9}{2}$ & $l=\dfrac{11}{2}$\\
\hline
$\lambda$ & 0.913149 & 3.97252 & 8.98335 & 15.9881 & 24.9907 & 35.9493\\
\hline
\multicolumn{7}{c}{$m_e=1, a=0.5, a\omega=10^{-3}$}\\
\hline
&$l=\dfrac{1}{2}$ & $l=\dfrac{3}{2}$ & $l=\dfrac{5}{2}$ & $l=\dfrac{7}{2}$ & $l=\dfrac{9}{2}$ & $l=\dfrac{11}{2}$\\
\hline
$\lambda$ & 0.48948 & 3.86191 & 8.91909 & 15.9409 & 24.9505 & 35.7377\\
\hline
\multicolumn{7}{c}{$m_e=1, a=0.9, a\omega=10^{-3}$}\\
\hline
&$l=\dfrac{1}{2}$ & $l=\dfrac{3}{2}$ & $l=\dfrac{5}{2}$ & $l=\dfrac{7}{2}$ & $l=\dfrac{9}{2}$ & $l=\dfrac{11}{2}$\\
\hline
$\lambda$ & 0.206311 & 3.78394 & 8.889633 & 15.9154 & 24.9218 & 35.5245\\
\hline
\multicolumn{7}{c}{$m_e=1, a=0.9999, a\omega=10^{-3}$}\\
\hline
&$l=\dfrac{1}{2}$ & $l=\dfrac{3}{2}$ & $l=\dfrac{5}{2}$ & $l=\dfrac{7}{2}$ & $l=\dfrac{9}{2}$ & $l=\dfrac{11}{2}$\\
\hline
$\lambda$ & 0.77948 & 3.77549 & 8.89233 & 15.9147 & 24.9179 & 35.4722\\
\hline
\end{tabular}
\caption{Values of the separation constant obtained with the method described in the text for $m_e=1$ and $a=\{0.1,0.5,0.9,0.9999\}$ for negative spin, $m=-\dfrac{1}{2}$ and $a\omega=\{0,10^{-3}\}$.}\label{tab:sep_con}
\end{table}

We can also calculate the approximate value of the separation constant for small values of $a\omega$ by expanding it and the function ${}sS_{lm}(\theta)$ in the following way:
\begin{align}
{}_sS_{lm}(\theta) &= {}_sY_{lm}(\theta) + a\omega\, S_1(\theta) + a^2\omega^2\,S_2(\theta) + O(a^3\omega^3)\\
\lambda &= \lambda_0 + a\omega\, \lambda_1 + a^2\omega^2\,\lambda_2 + O(a^3\omega^3),
\end{align}
where $\lambda_0$ is the value of the separation constant fo $a\omega=0$ calculated numerically as above. If we substitute the above expansions into \eqref{eq:angular} and separate the various orders of $a^n\omega^n$ we obtain the sought for terms, which are of course equal to the \emph{usual} expansion:
\begin{align}
\lambda_1 &= - 2m \left( 1+\dfrac{s^2}{l(l+1)} \right)\\
\lambda_2 &= H(l+1)-H(l) \qquad H(l) = \dfrac{2(l^2-m^2)(l^2-s^2)^2}{(2l-1)l^3(2l+1)}
\end{align}

We now turn to the radial equation.

\section{The radial Teukolky equation}
\label{sec:radial}

Similarly to the angular equation, also the radial Teukolky equation \eqref{eq:radial} is not in the form of an eigenvalue problem, and cannot be reformulated as such. The presence of a non-null fermion mass implies the presence of a third finite singularity in the differential equation. In \cite{mio}, we have found that we can write the general solution to the radial equation as:
\begin{equation}
{}_sR(x) = \exp(i \kappa \epsilon x)(-x)^{-s-\frac{i}{2}(\epsilon+\tau)}(1-x)^{\frac{i}{2}(\epsilon-\tau)} \, S_l(x),
\end{equation}
where $S_l(x)$ satisfies an equation of the form:
\begin{equation}\label{eq:o1}
\frac{d^2S_l}{dx^2} + \left( \frac{\gamma}{x} + \frac{\delta}{1-x}  + \frac{\eta}{A-x} \right) \frac{dS_l}{dx} + \frac{V(x)  S_l(x)}{x(1-x)(A-x)} = 0,
\end{equation}
where we have imposed:
\begin{equation}\label{eq:par}
\gamma=1-s -i \tau + i \epsilon \quad \delta=1+s-i\tau -i \epsilon \quad \eta= 1, \quad A=\frac{M-\kappa M-S_n}{2\kappa M} 
\end{equation}
with:
\begin{equation}
S_n=\left\{ \begin{array}{lr}
\dfrac{\sqrt{\lambda}}{2i M \tilde{m}_e s} & \text{for negative spin,}\\
\\
\dfrac{\sqrt{\lambda+1}}{2i M \tilde{m}_e s} & \text{for positive spin,}
\end{array} \right.
\end{equation}
while the potential $V(x)$ is given by:
\begin{equation}\label{eq:potential}
\begin{split}
V(x)&= (A-x)\,\Big[ \lambda+s(s+1)+\tau(i+\tau)-M^4\,\tilde{m}_e(1+\kappa(1-2x))^2 +\\
&+ \epsilon \, \Big( i\,\kappa(1-2s+2(-1+s+i \tau)) \Big) + \epsilon^2\,\Big( 1-2\kappa x \Big) \Big] + \frac{i\,\epsilon}{2} \, \Big( 1+2\kappa(x-1)x+s(2+4(-1+\kappa(x-1))x) \Big).
\end{split}
\end{equation}
In order to solve \eqref{eq:o1}, in \cite{mio}, we have noticed the similarity of this equation with an Heun differential equation. We have made use of this similarity by applying the formalism described in \cite{nist,expansion,expansion2} in order to expand the solution of \eqref{eq:o1} into a sum of Gauss' hypergeometric functions:
\begin{equation}
S_l(x) = \sum_n f^{(i)}_n\; {}_2F_1(\gamma+\delta+n-1,-n,\gamma,x) \qquad i=\{1,\dots,5\}.
\end{equation}
We have found that the five expansion parameters $f^{(i)}_n$ are the minimal solution to a seven-terms recurrence relation of the form:\footnote{For the mathematical treatment of higher order recurrece relations, see references \cite{recu1,recu2,recu3,recu4,recu5}.}
\begin{equation}\label{eq:recu}
\alpha^\nu_{0,n}a_{n}+\alpha^\nu_{1,n}a_{n+1}+\alpha^\nu_{2,n}a_{n+2}+\alpha^\nu_{3,n}a_{n+3}+\alpha^\nu_{4,n}a_{n+4}+\alpha^\nu_{5,n}a_{n+5}+\alpha^\nu_{6,n}a_{n+6}=0
\end{equation}
where the $\alpha_{i,n}$, $i=\{0,\dots,6\}$ are listed in the appendix. 

Similarly to the usual case, in \cite{mio}, we have introduced a \emph{renormalized angular momentum} $\nu$; we do this here too, by introducing the expansion:
\begin{equation}\label{eq:nu_exp}
\nu = \nu_0 + \epsilon^2\, \nu_2 + O(\epsilon^3),
\end{equation}
where $\nu_0$ is the renormalized angular momentum for $\epsilon = 0$, which must be calculated numerically. We notice, however, that differently from the usual case, there are now 5 different $\nu^{(i)}$, one for each solution to the recurrence relation. Similarly to the usual case, we introduce the ten operators:
\begin{equation}
R^{(i)}_n = \dfrac{f^{(i)}_n}{f^{(i)}_{n-1}} \qquad L^{(i)}_n = \dfrac{f^{(i)}_n}{f^{(i)}_{n+1}} \qquad i=\{1,\dots,5\},
\end{equation}
each pair of which satisfy the relation:
\begin{equation}\label{eq:cond}
R^{(i)}_n L^{(i)}_{n-1} = 1.
\end{equation}
These are the five equations which must be solved numerically in order to find the various $\nu_0^{(i)}$. However, we have found that only $R^{(i)}_n L^{(i)}_{n-1} = 1$ for $i=2$ has solutions, which are reported in table \ref{tab:valnu} for $\tilde{m}_e=100$ (obtained with $m_e=1, M=10^{-2}$) and $a=\{0.1,0.5,0.9,0.9999\}$ for negative spin and $m=\dfrac{1}{2}$ as an example.

\begin{table}
\centering
\begin{tabular}{cccc}
\multicolumn{4}{c}{$a=0.1, \tilde{m}_e=100, m=\dfrac{1}{2}$}\\
\hline
& $l=\dfrac{1}{2}$& $l=\dfrac{3}{2}$& $l=\dfrac{5}{2}$\\
\hline
$\nu$ & $-0.01215 + i 0.08485$ & $1.49764$ & $2.49749$\\
\hline
\multicolumn{4}{c}{$a=0.5, \tilde{m}_e=100, m=\dfrac{1}{2}$}\\
\hline
& $l=\dfrac{1}{2}$& $l=\dfrac{3}{2}$& $l=\dfrac{5}{2}$\\
\hline
$\nu$ & $0.518694 + i 0.0.000248$ & $1.52693 - i 0.001025$ & $2.5024 - i 0.000104$\\
\hline
\multicolumn{4}{c}{$a=0.9, \tilde{m}_e=100, m=\dfrac{1}{2}$}\\
\hline
& $l=\dfrac{1}{2}$& $l=\dfrac{3}{2}$& $l=\dfrac{5}{2}$\\
\hline
$\nu$ & $0.51398 + i 0.00006$ & $1.63086 - i 0.005417$ & $2.51666 - i 0.000970$\\
\hline
\multicolumn{4}{c}{$a=0.9999, \tilde{m}_e=100, m=\dfrac{1}{2}$}\\
\hline
& $l=\dfrac{1}{2}$& $l=\dfrac{3}{2}$& $l=\dfrac{5}{2}$\\
\hline
$\nu$ & $0.515553 + i 0.000072$ & $1.55873 - i0.001589$ & $2.50599 - i 0.000192$\\
\hline
\end{tabular}
\caption{Values of $\nu=\nu_0$ calculated from the numerical solution of \eqref{eq:cond} with $i=2$, since the others have no solutions. We have used the following values for the parameters: $\tilde{m}_e=100$ (obtained with $m_e=1, M=10^{-2}$) and $a=\{0.1,0.5,0.9,0.9999\}$ for negative spin and $m=\dfrac{1}{2}$.}\label{tab:valnu}
\end{table}

For small $\epsilon$, we can also find analytically the correction $\nu_2$ in \eqref{eq:nu_exp} by substituting the expansion \eqref{eq:nu_exp} into the recurrence relation and equating to zero the coefficient at $O(\epsilon^2)$. We have found in this way the correction, but the expression is very large, so we do not report it here, but we make available the Mathematica notebooks used for our calculation and the files containing the expressions.

\section{Conclusions}
\label{sec:concl}
In this work we have concluded the treatment of the Teukolsky equations for massive fermion we have started in \cite{mio}. There we have treated the cases of small $\tilde{m}_e = m_e\, M^{-1}$, here we have treated the cases of large $\tilde{m}_e$. We have presented a numerical method for the calculation of the separation constant $\lambda$ inspired to the spectral method presented in \cite{huges1,huges2} and also analytical corrections to $\lambda$ and $\nu$ in the case of small $a\omega$.

\appendix
\section{Recurrence relations coefficients}
\label{app:uno}
In this appendix we report the full expression of the coefficients of the recurrence relations \eqref{eq:recu}.
\begin{equation*}
\alpha^\nu_{0,n+\nu} = T_3\, N^{(1)}_{n-2}\, N^{(1)}_{n-1}\, N^{(1)}_{n},
\end{equation*}
\begin{equation*}
\begin{split}
\alpha^\nu_{1,n+\nu} &= T_2+T_3 \, \Big( N^{(2)}_{n-2} + N^{(2)}_{n-1} + N^{(2)}_n \Big)\, N^{(1)}_{n-1}\,N^{(1)}_n,
\end{split}
\end{equation*}
\begin{equation*}
\begin{split}
\alpha^\nu_{2,n+\nu} &= N^{(1)}_n\, \Big[ T_1 + n\, \Big( T_{11} + n\, T_{12} \Big) + T_3\,(N^{(2)}_{n-1})^2 + T_2\, N^{(2)}_n + T_3\, (N^{(2)}_n)^2 +\\
&+ N^{(2)}_{n-1} \,\Big( T_2 +T_3\, N^{(2)}_n \Big) + T_3 \Big( N^{(3)}_{n-2} N^{(1)}_{n-1} + N^{(3)}_{n-1} N^{(1)}_{n} + N^{(3)}_{n} N^{(1)}_{n+1} \Big) \Big] + ND^{(1)}_n,
\end{split}
\end{equation*}
\begin{equation*}
\begin{split}
\alpha^\nu_{3,n+\nu} &= T_0 + n\,\Big( T_{01} + n  T_{02} \Big) + T_2\, (N^{(2)}_n)^2 + T_3\,(N^{(2)}_n)^3 + ND^{(1)}_n + \Big( T_2 + T_3\, N^{(2)}_{n-1} \Big) \, N^{(3)}_{n-1}\,N^{(1)}_n +\\
&+ \Big( T_2 + T_3 \,N^{(2)}_{n+1} \Big) \, N^{(3)}_n\, N^{(1)}_{n+1} + N^{(2)}_n, \Big[ T_1 + n \, \Big( T_1 + n \, T_{11} + n^2\, T_{12} + 2 T_3\,N^{(3)}_{n-1}\,N^{(1)}_n + 2 \, T_3\, N^{(3)}_n\, N^{(1)}_{n+1} \Big)  \Big],
\end{split}
\end{equation*}
\begin{equation*}
\begin{split}
\alpha^\nu_{4,n+\nu} &= ND^{(3)}_n + N^{(3)}_n\, \Big[ T_1 + n\, T_{11} + n^2\, T_{12} + T_3\, (N^{(2)}_n)^2 + T_2 \, N^{(2)}_{n+1} + T_3\, (N^{(2)}_{n+1})^2 + N^{(2)}\, \Big( T_2 + T_3 \, N^{(2)}_{n+1} \Big) +\\
&+ T_3\, \Big( N^{(3)}_{n-1}\,N^{(1)}_n + N^{(3)}_n\,N^{(1)}_{n+1} + N^{(3)}_{n+1}\,N^{(1)}_{n+2} \Big) \Big],
\end{split}
\end{equation*}
\begin{equation*}
\begin{split}
\alpha^\nu_{5,n+\nu} &= \Big[ T_2 +T_3\, \Big( N^{(2)}_n + N^{(2)}_{n+1} + N^{(2)}_{n+2} \Big)  \Big] \, N^{(3)}_n\, N^{(3)}_{n+1},
\end{split}
\end{equation*}
\begin{equation*}
\begin{split}
\alpha^\nu_{6,n+\nu} &= T_3\, N^{(3)}_n\, N^{(3)}_{n+1}\, N^{(3)}_{n+2}.
\end{split}
\end{equation*}

The expressions for $ND^{(i)}_n$ and $N^{(i)}_n$ are given by
\begin{subequations} \label{eq:def_ND}
\begin{equation}
ND^{(1)}_{n-1}=-\frac{(\mu-n)(\sigma+n)(\sigma+n-\gamma)}{\sigma-\mu-1+2n)(\sigma-\mu+2n)},
\end{equation}
\begin{equation}
ND^{(2)}_{n}=\frac{(1-2\gamma+\sigma+\mu)(\mu-n)(\sigma+n)}{(\sigma-1-\mu+2n)(\sigma+1-\mu+2n)},
\end{equation}
\begin{equation}
ND^{(3)}_{n+1}=\frac{(\mu-n)(\sigma+n)(\gamma-\mu+n)}{(\sigma-\mu+2n)(1+\sigma-\mu+2n)}.
\end{equation}
\end{subequations}
\begin{subequations} \label{eq:def_N}
\begin{equation}
N^{(1)}_{n-1}=\frac{(\mu-n)(\sigma-\gamma+n)}{(\sigma+2n-\mu-1)(\sigma+2n-\mu)},
\end{equation}
\begin{equation}
N^{(2)}_{n}=\frac{(\mu-1)\,\gamma+2n\,(n-\mu)+(\sigma+n)\,(\gamma-2\mu+2n)}{(\sigma+2n-1-\mu)(1+2n+\sigma-\mu)},
\end{equation}
\begin{equation}
N^{(3)}_{n+1}=\frac{(\mu-\gamma-n)(\sigma+n)}{(\sigma-\mu+2n)(1+\sigma-\mu+2n)},
\end{equation}
\end{subequations}
while the definitions of the coefficients appearing in the above expressions are given below:
\begin{equation*}
T_0 = \dfrac{1}{4}\, \Big[ 1+\, \gamma(4+\gamma) - (\delta-2) + 4\, A\, \Big( \lambda - \sigma \mu-(1+\kappa)^2M^4 \tilde{m}_e^2 + (1+\gamma)\delta + i\, \epsilon\, (\delta-1-\gamma + \kappa, (\delta+\gamma-1)) \Big) \Big],
\end{equation*}
\begin{equation*}
T_{01} = A\,( \sigma-\mu ),
\end{equation*}
\begin{equation*}
T_{02} = A,
\end{equation*}
\begin{equation*}
T_1 = 4\, A\, \epsilon^2\,\kappa-\lambda + \sigma\, \mu + (\kappa+1)(8A\,\kappa-\kappa-1) M^4\, \tilde{m}_e^2 + i \, \epsilon\, \Big[ 2\kappa(-1+A\,\gamma)-1-2\gamma \Big] - \delta (\gamma+1),
\end{equation*}
\begin{equation*}
T_{11} = -\sigma+\mu
\end{equation*}
\begin{equation*}
T_{12} = -1,
\end{equation*}
\begin{equation*}
T_2 = \kappa\, \Big[ -4\, \epsilon^2-4(1+\kappa+A\,\kappa)\,M^4\, \tilde{m}_e^2 - i\, \epsilon\,(\gamma-\delta-1) \Big],
\end{equation*}
\begin{equation*}
T_3 = 4\kappa^2\,M^4\, \tilde{m}_e^2.
\end{equation*}

We also remind the definition of $A$:
\begin{equation*}
A = \left\{\begin{array}{lr}
\dfrac{1}{2} + \dfrac{m\,s\,M^2\,\tilde{m}_e+i\,\sqrt{\lambda}}{2m\,s\,\kappa\,M^2\,\tilde{m}_e} & s<0,\\
\\
\dfrac{1}{2} + \dfrac{m\,s\,M^2\,\tilde{m}_e+i\,\sqrt{\lambda+1}}{2m\,s\,\kappa\,M^2\,\tilde{m}_e} & s>0.\\
\end{array}\right.
\end{equation*}


\begin{thebibliography}{}
\bibitem{mio} M. Villani, arXiv:2310.13645 [gr-qc];
\bibitem{lrr} M. Sasaki \& H. Tagoshi, \emph{Living Rev. Relativity} \textbf{6} (2003), 6;
\bibitem{libro} J. A. H. Futterman, F. A. Handler, R. A. Matzner, \emph{Scattering from black holes}, Cambridge University Press, Cambridge, (2009);
\bibitem{Ferm} S. R. Dolan et al. , \emph{Phys. Rev. D} \textbf{74} (2006) 064005;
\bibitem{SD} S. R. Dolan, \emph{Class. Quantum Grav.} \textbf{25} (2008) 235002;
\bibitem{huges1} S. A. Huges, \emph{Phys. Rev. D} \textbf{61} (2000) 084004;
\bibitem{huges2} S. A. Huges, \emph{Phys. Rev. D} \textbf{63} (2001) 049902;
\bibitem{SWSA} R. A. Breuer et al., \emph{Proc. Royal Soc. London A} \textbf{358} (1977) 71;
\bibitem{nist} F. W. J. Olver, D. Lozier, R. Boisvert, C. Clark, \emph{NIST Handbook of Mathematical Function}, Cambridge University Press, New York (2010);
\bibitem{expansion} A. Erd\'elyi, \emph{The Quarterly Journal of Mathematics} \textbf{15} (1944) 62;
\bibitem{expansion2} A. Erd\'elyi, \emph{The Quarterly Journal of Mathematics}, \textbf{13} (1942) 107;
\bibitem{recu1} M. C. de Bruin, \emph{Journal of Approximation Theory}, \textbf{24} (1978) 177;
\bibitem{recu2} P. Van der Cruyssen, \emph{Computing} \textbf{22} (1979) 269;
\bibitem{recu3} P. Van der Cruyssen, \emph{International Journal of Computer Mathematics}, \textbf{10} (1982) 295;
\bibitem{recu4} M. C. de Bruin, \emph{Journal of  Computational and Applied Mathematics}, \textbf{9} (1983) 271;
\bibitem{recu5} P. Levrie, \emph{Applied Numerical Mathematics}, \textbf{8} (1991) 225.
\end{thebibliography}
\end{document}